%% file: main.tex
\documentclass{article}

\usepackage[colorlinks]{hyperref}

\usepackage{setspace}
\usepackage{multicol}

\usepackage{amsmath,amssymb,color}
\usepackage{graphicx}
\usepackage{braket}
\usepackage{cite}
\usepackage[short]{optidef}
\usepackage{bm}

\usepackage{enumerate}

\usepackage[preprint, nonatbib]{neurips_data_2023}

\usepackage[utf8]{inputenc} % allow utf-8 input
\usepackage[T1]{fontenc}    % use 8-bit T1 fonts
\usepackage{hyperref}       % hyperlinks
\usepackage{url}            % simple URL typesetting
\usepackage{booktabs}       % professional-quality tables
\usepackage{amsfonts}       % blackboard math symbols
\usepackage{nicefrac}       % compact symbols for 1/2, etc.
\usepackage{microtype}      % microtypography
\usepackage{xcolor}         % colors
\usepackage{qcircuit}       % quantum circuit
\usepackage{braket}         % braket icon
\usepackage{algorithm,algorithmic}
\usepackage{listings}
\usepackage{multibib}
\newcites{supp}{Supplementary References}

\title{MNISQ: \\ 
A Large-Scale Quantum Circuit Dataset for Machine Learning on/for Quantum Computers in the NISQ era}
\author{%
   Leonardo Placidi \\
   Osaka University - QIQB \\
   Osaka, Japan\\
   \texttt{u770335b@ecs.osaka-u.ac.jp}
   \AND
   Ryuichiro Hataya \\
   RIKEN \\
   Tokyo, Japan\\
   \texttt{ryuichiro.hataya@riken.jp} \\
   \And
   Toshio Mori \\
   Osaka University - QIQB - RIKEN\\
   Osaka - Wako Saitama, Japan\\
   \texttt{t.mori.qiqb@osaka-u.ac.jp} \\
   \And
   Koki Aoyama \\
   Osaka University \\
   Osaka, Japan \\
   \texttt{k-aoyama@ist.osaka-u.ac.jp} \\
   \And
   Hayata Morisaki \\
   Osaka University \\
   Osaka, Japan  \\
   \texttt{u748119d@ecs.osaka-u.ac.jp} \\
   \AND
   Kosuke Mitarai \\
   Osaka University - QIQB \\
   Osaka, Japan
   \\
   \texttt{mitarai.kosuke.es@osaka-u.ac.jp} \\\And
   Keisuke Fujii \\
   Osaka University - QIQB - RIKEN \\
   Osaka - Wako Saitama, Japan\\
   \texttt{fujii.keisuke.es@osaka-u.ac.jp} \\
}

\begin{document}

\maketitle
\input{arXiv_sections/abstract}
\input{arXiv_sections/introduction}

\input{arXiv_sections/preliminaries}
\input{arXiv_sections/AQCE}
\section{Experimental results}
\label{sec:experiments}
%In this section, we present our experimental results on circuit classification using both quantum and classical models. 
\input{arXiv_sections/QML}

\input{arXiv_sections/ML}
\input{arXiv_sections/conclusion}

\begin{ack}
This work is supported by MEXT Quantum Leap Flagship Program (MEXT Q-LEAP) Grant No. JPMXS0118067394 and JPMXS0120319794, and JST COI- NEXT Grant No. JPMJPF2014. We finally acknowledge the use of LLMs (\cite{chatgpt}) for improving our drafts.
\end{ack}
\bibliographystyle{unsrt}
\bibliography{arxivMNISQ}
\newpage
\appendix
\input{arXiv_supplementary/appendix}
\newpage
\input{arXiv_supplementary/data_gen}
\newpage
\section{Machine Learning models and details}
\input{arXiv_supplementary/S4}
\input{arXiv_supplementary/TF}
\input{arXiv_supplementary/LSTM}

\input{arXiv_supplementary/models}
\input{arXiv_supplementary/license}
\input{arXiv_supplementary/metadata}
\input{arXiv_supplementary/statement}
\input{arXiv_supplementary/url}
\bibliographystylesupp{unsrt}
\bibliographysupp{arxivMNISQ}
%\textbf{Authors of this file}: Leonardo Placidi, Ryuichiro Hataya, Toshio Mori, Koki Aoyama, Hayata Morisaki, Kosuke Mitarai and Keisuke Fujii.
%%%%%%%%%%%%%%%%%%%%%%%%%%%%%%%%%%%%%%%%%%%%%%%%%%%%%%%%%%%%

\end{document}

%% file: arXiv_sections/abstract.tex
\begin{abstract}
We introduce the first large-scale dataset, MNISQ, for both the Quantum and the Classical Machine Learning community during the Noisy Intermediate-Scale Quantum era. MNISQ consists of 4,950,000 data points organized in 9 subdatasets. Building our dataset from the quantum encoding of classical information (e.g., MNIST dataset), we deliver a dataset in a dual form: in quantum form, as circuits, and in classical form, as quantum circuit descriptions (quantum programming language, QASM).
In fact, also the Machine Learning research related to quantum computers undertakes a dual challenge: enhancing machine learning exploiting the power of quantum computers, while also leveraging state-of-the-art classical machine learning methodologies to help the advancement of quantum computing.
Therefore, we perform circuit classification on our dataset, tackling the task with both quantum and classical models.
In the quantum endeavor, we test our circuit dataset with Quantum Kernel methods, and we show excellent results up to $97\%$ accuracy. In the classical world, the underlying quantum mechanical structures within the quantum circuit data are not trivial. Nevertheless, we test our dataset on three classical models: Structured State Space sequence model (S4), Transformer and LSTM. In particular, the S4 model applied on the tokenized QASM sequences reaches an impressive $77\%$ accuracy. These findings illustrate that quantum circuit-related datasets are likely to be quantum advantageous, but also that state-of-the-art machine learning methodologies can competently classify and recognize quantum circuits. We finally entrust the quantum and classical machine learning community the fundamental challenge to build more quantum-classical datasets like ours and to build future benchmarks from our experiments. The dataset is accessible on \href{https://github.com/FujiiLabCollaboration/MNISQ-quantum-circuit-dataset/tree/main}{GitHub} and its circuits are easily run in qulacs or qiskit.
\end{abstract}

%% file: arXiv_sections/introduction.tex
\section{Introduction}
\label{sec:intro}
\subsection{Background}
The advent of quantum computers has garnered significant attention across various scientific fields. By harnessing the principles of quantum mechanics, these computers are capable of performing calculations that are beyond the capabilities of classical computers~\cite{nielsen_chuang}. 
While classical computers operate using bits that can be either $0$ or $1$, quantum computers use qubits that can take a superposed state of $0$ and $1$. This quantum property enables calculations much faster than those on classical computers.  As a result, quantum computers excel at solving specific problems, such as prime factorization~\cite{shor1994algorithms}, simulations of quantum many-body systems~\cite{lloyd1996universal}, and linear system solvers~\cite{harrow2009quantum}.

With the realization of quantum computers comprising $50$ to $100$ qubits, the field has entered the era of quantum computational supremacy~\cite{arute2019quantum,wu2021strong,zhu2022quantum,morvan2023phase}. In this era, even the most powerful supercomputers struggle to simulate the behavior of quantum computers.  However, this argument is based on a benchmark task known as random quantum circuit sampling~\cite{arute2019quantum}, and the usefulness of quantum computers in more practical and meaningful tasks is yet to be fully realized.
Developing algorithms that effectively leverage the power of current quantum computers to solve real-world problems remains a significant challenge. The current generation of quantum computers, known as NISQ (Noisy Intermediate-Scale Quantum) computers \cite{preskill2018quantum}, are not yet fully error-corrected and rely on small-to-medium-scale devices that are subject to noise.
Among the most promising applications of NISQ devices are variational quantum algorithms \cite{cerezo2021variational}, particularly in the context of quantum machine learning (QML) \cite{biamonte2017quantum, cerezo2021variational, cerezo2022challenges}. These algorithms have attracted considerable attention as they are employed to address various problems like classification, regression, and generative modeling. 
QML is currently at a phase similar to the early days of classical machine learning, with numerous applications being proposed and explored.
However, when using NISQ computers, QML is primarily capable of solving only “toy” problems \cite{cerezo2022challenges}, and it cannot yet directly compete with state-of-the-art classical machine learning models, such as ChatGPT \cite{chatgpt}.

\subsection{Motivation}
The motivation behind this study is to establish datasets that demonstrate the advantage of Quantum Machine Learning (QML) and compare its performance against conventional classical machine learning approaches. While it has been theoretically shown that QML can be advantageous for certain tasks, there is a need to explore practical and experimentally friendly settings to showcase the potential of QML. 

One approach is to work with quantum data obtained from quantum systems directly, without converting them to classical formats, as it has been shown to be more efficient for certain learning tasks~\cite{Huang2022}. However, this approach presents challenges in transferring quantum states obtained from physical experiments to quantum computers, making it experimentally challenging. Another approach is to define artificial machine learning tasks that are solvable only by quantum computers, such as the discrete logarithm problem~\cite{liu2021rigorous}. While this demonstrates a rigorous advantage of QML, it is in a highly artificial setting and may not directly translate to advantages in practical problems.

To address these limitations and establish a more practical and experimentally friendly setting, it is crucial to create datasets where the advantage of QML can be expected and compare its performance against classical approaches. One such dataset, called NTangle~\cite{schatzki2021entangled}, has taken a step towards this direction by using quantum states labeled by their level of entanglement. However, applying classical machine learning to this dataset and comparing it with QML approaches is challenging due to its quantum nature.

To overcome this challenge, a dataset described in a quantum-related classical language called Quantum Assembly Language (QASM)\cite{cross2022openqasm} has been generated using real-world quantum computing tasks\cite{nakayama2023vqe}. Although this approach is interesting, the dataset size is limited to only $1800$ elements due to the computational cost of data generation. Additionally, such a small dataset size may not be sufficient to demonstrate that classical machine learning techniques are unable to learn the dataset, as modern machine learning research typically utilizes much larger datasets. Existing quantum-related datasets, such as PennyLane~\cite{bergholm2022pennylane}, also suffer from limitations such as small sizes and the inability to be used for both quantum and classical machine learning.

\begin{figure}[H]
  \centering
  \includegraphics[width=0.95\linewidth]{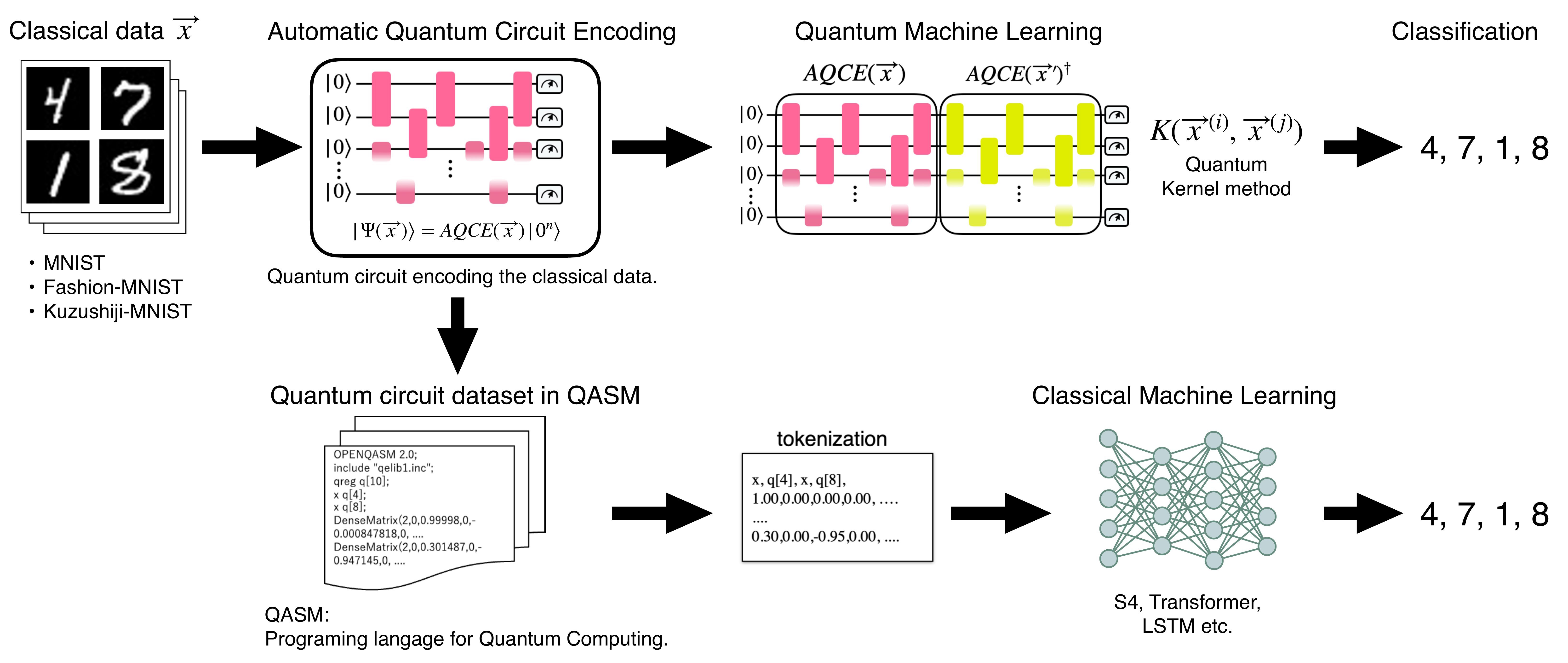}
  \caption{Summary of the workflow for this study. The AQCE algorithm generates a quantum circuit that embeds classical data $\vec{x}$ into a quantum state $|\Psi \rangle = AQCE(\vec{x})|0^n\rangle$ and delivers a dataset in dual form. Quantum machine learning (quantum kernel method) and classical machine learning (LSTM, S4, Transformer) are applied to classify the data encoded in the quantum circuits or qasm files.}
  \label{fig_MNISQ}
\end{figure}

In summary, the motivation of this study is to address the limitations of existing datasets and establish a dataset, called MNISQ, that allows for the comparison of QML and classical machine learning approaches. The MNISQ dataset is generated using the AQCE algorithm~\cite{shirakawa2021automatic}, which generates quantum circuits that encode classical data into quantum states. This dataset is provided in a dual form that can be utilized for both quantum and classical machine learning. The workflow of this study is summarized in Figure \ref{fig_MNISQ}.

\subsection{Approach}
In this work, we introduce a novel large-scale quantum machine learning dataset, which consists of hundreds of thousands of labeled quantum circuits. Each circuit is accompanied by a classical description in the QASM programming language. This dataset represents the largest collection of quantum circuits to date (4,950,000 data points organized in $9$ subdatasets) and aims to address the limitations and challenges faced by the quantum machine learning (QML) community.

Our objectives are twofold. First, we utilize this dataset to seek for a quantum advantage in meaningful tasks by executing it on relatively small-scale quantum computers. Second, we leverage the vast quantum circuit dataset to explore the classification problem of quantum circuits using state-of-the-art classical machine learning techniques, thereby challenging the extent to which classical methods can recognize quantum circuits.

To create the quantum circuit dataset MNISQ, we employ the automatic quantum circuit encoding (AQCE) method, which embeds the MNIST dataset into the complex amplitudes of quantum states \cite{shirakawa2021automatic}. This systematic approach allows us to generate a massive number of quantum circuits, with each circuit encoding a digit number in its quantum state representation when executed on a quantum computer. MNISQ is compatible with modern quantum hardware (10-qubits), including NISQ devices, despite the presence of noise and deviations from ideal conditions. Future work could involve experimental demonstrations of quantum machine learning using the MNISQ dataset, leveraging the robustness of the quantum kernel method as demonstrated in previous experiments~\cite{Havl_ek_2019, kusumoto2019experimental}.

For classification, we employ the Quantum Support Vector Machine (QSVM) algorithm \cite{Havl_ek_2019, Mengoni2019KernelMI}, a powerful QML algorithm. Our experiments demonstrate outstanding performance, achieving up to $97.91\%$ accuracy on the MNIST dataset, with results ranging from $81\%$ to $97\%$ accuracy.

Furthermore, we investigate the learnability of the quantum circuit dataset by classical machine learning methods. Although the task is significantly more challenging due to the nature of the text data, we successfully train models such as Structured Search Space (S4)~\cite{gu2022efficiently}, Transformers~\cite{vaswani2017attention}, and LSTM~\cite{hochreiter1997long}. Surprisingly, our experiments reveal that the S4 model performs the best, achieving $77.78\%$ accuracy. These results indicate that while our dataset clearly is likely quantum advantageous, it is also well learnable by classical machine learning models.

The significance of datasets in the history of machine learning cannot be overstated, as they have played a pivotal role in the field's development and progress (e.g., think of the impact of mnist or imagenet\cite{mnist, imagenet}). Similarly, the establishment of quantum-related datasets is crucial for the healthy growth of QML research. Additionally, our work introduces a new field in classical machine learning, namely, learning quantum circuits. By leveraging the robust capabilities of classical machine learning, we aim to design and develop efficient quantum algorithms.

In conclusion, our research presents exciting opportunities for researchers and practitioners to explore and exploit the potential advantages of quantum computing in machine learning and vice versa. This work paves the way for breakthroughs and discoveries that may shape the future of computing and artificial intelligence.
\subsection{Contributions}

\begin{itemize}
\item We introduce a large-scale quantum dataset in the form of quantum circuits and QASM, facilitating research in quantum machine learning and classical machine learning. The dataset is easily accessible and, given its QASM formalism, easily used in qulacs or qiskit.
\item We demonstrate the outstanding performance of a quantum support vector machine on the dataset, suggesting potential quantum advantage.
\item We explore the classification of circuits using classical models (S4, Transformers, LSTM), revealing the solvability of the problem by classical machine learning.
\item We highlight the remarkable capability (and the implications) of classical machine learning in comprehending quantum circuit computations without prior knowledge of quantum mechanics.
\end{itemize}

%% file: arXiv_sections/preliminaries.tex
\section{Preliminaries}
In classical computation, information is represented via binary bits; however, in the realm of quantum computing, these superposition states can be exploited through the principle of superposition inherent in quantum mechanics. 
Such superposed states assign complex numbers, referred to as "complex probability amplitudes", to each potential state of $n$ classical binary bits, thereby expressing them as a $2^n$-dimensional complex vector. 
Quantum computers execute computation by 
transforming these $2^n$-dimensional complex vectors 
via physical allowed operations, i.e., unitary matrices.
Thereby, quantum computers provide quantum speedup for solving certain class of problems that are intractable on classical computers, such as the prime factorization. 
Machine learning is one of the fields in which quantum computers are expected to have an advantage, especially through the high-dimensional vector space that quantum computers represent.
Additionally, the realization of quantum computers presents substantial challenges both experimentally and in terms of software, suggesting the potential direction of employing existing machine learning for the benefit of quantum computers.
The interdisciplinary field between quantum computing and machine learning is now driven by two major challenges:
enhancing machine learning exploiting the power of quantum computers, while also leveraging state-of-the-art classical machine learning methodologies to help the advancement of quantum computing.\\

For readers of any background, we included a detailed introduction to quantum computing in the supplementary materials.

%% file: arXiv_sections/AQCE.tex
\section{MNISQ Dataset: Construction and Characteristics}
\label{sec:aqce}

\begin{figure}[H]
  \centering
  \includegraphics[width=0.8\linewidth]{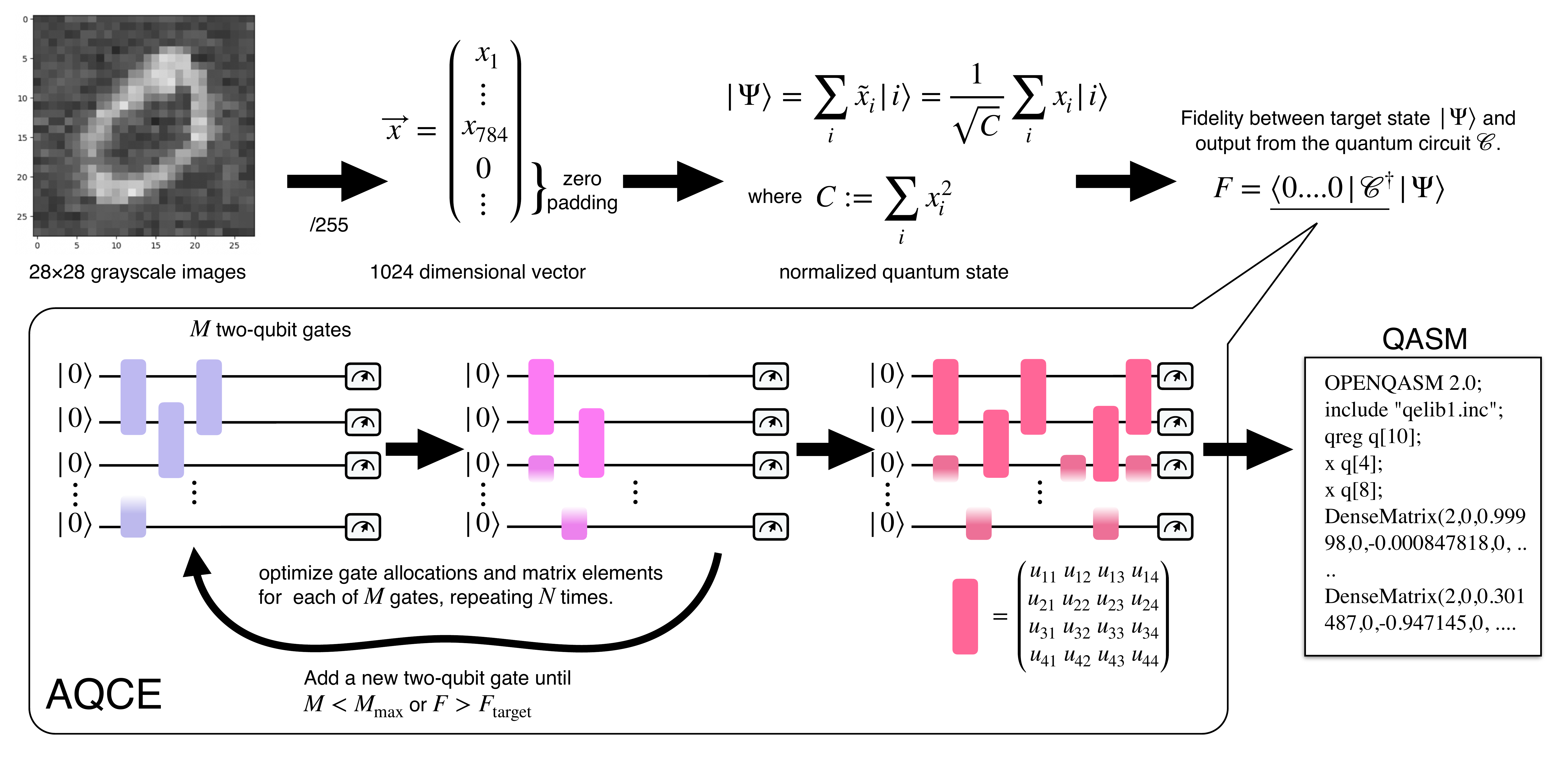}
  \caption{Schematic diagram of how to embed classical data into the probability amplitude of a quantum state the using the AQCE~\cite{shirakawa2021automatic} algorithm.}
  \label{fig_AQCE}
\end{figure}
To enable quantum machine learning on classical data like MNIST, it is crucial to transform the classical data into quantum states. In this study, we utilize the Automatic Quantum Circuit Encoding (AQCE) approach \cite{shirakawa2021automatic} to achieve this embedding process. Figure~\ref{fig_AQCE} illustrates the procedure.

\subsection{Quantum Encoding of Classical Information (AQCE method)}
\label{subsec:aqce}
Note that, for readers of any background, we provide a detailed introduction to quantum computing in the supplementary materials.

Given a classical data $\vec{x}$ as $d$-dimensional vector,
AQCE generates a quantum circuit $\mathcal{C}$ that encode the classical data $\vec{x}$
into the complex amplitudes of the output quantum state:
\begin{align}
|\Psi \rangle := \sum _{i \in \{{0,1}\}^n} \tilde x_i |i \rangle ,
\end{align}
where the number of qubits $n$ is chosen to be $n:=\lceil \log_2(d) \rceil$ and 
complex amplitudes $\tilde x_i$ is normalized appropriately so that 
$\sum _i |\tilde x_i|^2 = 1 $.
If $2^n > d$, for those indices that do not have corresponding elements in $\vec{x}$, $\tilde x_i$ is set to be zero.
AQCE constructs such a quantum circuit $\mathcal{C}$ generating any given quantum state $\ket{\Psi}$ by optimizing the configuration and parameters of the quantum gates in $\mathcal{C}$
by maximizing the absolute value of the following overlap, which we call fidelity here:
\begin{equation}
F =|\bra{0} \mathcal{C}^\dagger \ket{\Psi}|.
\end{equation}
The AQCE iteration ends when the fidelity exceeds the target fidelity $F_{\rm target}$ or the maximum number of gates $M_{\rm max}$ initially set for $\mathcal{C}$. The resulting quantum circuit generated by the AQCE procedure from classical data $\vec{x}$ 
will be referred to as the $AQCE(\vec{x})$ circuit.

The above AQCE procedure is intended to be performed by simulating a quantum computation on a classical computer. 
Although this can only be done on the order of tens of qubits, we believe that the dimension of the classical data that can be amplitude embedded is sufficient, for example, $10^9$ for a classical simulation of $30$ qubits. Also, the embedding operation with AQCE may cause a computational bottleneck, but once the quantum circuit data is generated in this format (like JPG format for images), it can be used efficiently thereafter.
For a more detailed description of AQCE, please refer to the supplementary material paragraph on the AQCE method.

\subsection{Dataset Construction}
Using a cluster machine, we employed the AQCE method to encode the standard machine learning datasets MNIST \cite{mnist}, Fashion-MNIST \cite{xiao2017fashionmnist}, and Kuzushiji-MNIST \cite{kuzushijimnist}. These datasets were obtained from OpenML \cite{OpenML2013}.

For each dataset, we generated 60,000 training samples and 10,000 test samples. Additionally, to train classical machine learning models, we augmented our quantum dataset to include 480,000 samples. The data augmentation involved applying various transformations to the original images, such as rotation (random angle of ±$50$ degrees), rotation followed by cropping, and rotation-cropping combined with shifting. This resulted in a total of 480,000 training samples for each of the three datasets. We preserved the original test data without augmentation to ensure a direct comparison between the predictions of the quantum and classical models.

Since the AQCE method allows adjusting the fidelity of the quantum state by specifying parameters, we created three variations for each dataset. These variations correspond to different generation procedures with a minimum fidelity of $80\%$, $90\%$, and $95\%$ for the generated quantum circuit dataset. Higher fidelity values require a larger number of quantum gates in the circuit. 

\subsection{Dataset Accessibility}\label{subs:dataconstr}

The datasets we have created are easily accessible through the MNISQ library. To install the library from PyPI, use the following command: \verb|pip install mnisq|. Alternatively, the datasets can be accessed directly from the URLs provided below.

The MNISQ library automatically downloads predefined quantum circuits for Qulacs (Quantum Circuit Simulator), which are ready for use. When executing a circuit starting from the initial zero state, one of the images from the dataset is embedded in the quantum state.

\textbf{Note}: The QASM files come in two different forms:

\begin{itemize}
    \item \textit{QASM with Dense() formalism}: These files can be used in Qulacs but not in Qiskit or other platforms due to the proprietary Dense() operator.
    \item \textit{Base QASM formalism}: \textit{These files can be run in different platforms, including Qiskit~\cite{Qiskit}}. A tutorial on how to run them in Qiskit can be found on our GitHub page at the following link: \href{https://github.com/FujiiLabCollaboration/MNISQ-quantum-circuit-dataset/blob/main/doc/source/notebooks/qiskit_quickstart.ipynb}{tutorial link}. Please note that to access these files, they begin with the prefix \textit{"base\_.."}.
\end{itemize}

The datasets can be found at the following URL:

\textit{https://qulacs-quantum-datasets.s3.us-west-1.amazonaws.com/[data]\_[type]\_[fidelity].zip}

Here are the parameters for the URL:

\begin{itemize}
    \item \textbf{data}:
    \begin{enumerate}
        \item \textbf{"train\_orig"}: 60,000 original encoded training data QASM files with \textit{Dense()} formalism.
        \item \textbf{"base\_train"}: Same as above, but the QASM files do not include the \textit{Dense()} operator. Instead, they use a gate conversion and can be run on Qiskit and other platforms.
        \item \textbf{"train"}: "train\_orig" augmented to 480,000 training data for each subdataset using \textit{Dense()} formalism.
        \item \textbf{"test"}: 10,000 test elements from the original encoding. QASM files with \textit{Dense()} formalism.
        \item \textbf{"base\_test"}: Same as above, but the QASM files do not include the \textit{Dense()} operator. Instead, they use a gate conversion and can be run on Qiskit and other platforms.
    \end{enumerate}
    \item \textbf{type}:
    \begin{enumerate}
        \item \textbf{"mnist\_784"}: MNIST dataset.
        \item \textbf{"Fashion-MNIST"}
        \item \textbf{"Kuzushiji-MNIST"}
    \end{enumerate}
    \item \textbf{fidelity}:
    \begin{enumerate}
        \item \textbf{"f80"}: Fidelity greater than or equal to 80%.
        \item \textbf{"f90"}: Fidelity greater than or equal to 90%.
        \item \textbf{"f95"}: Fidelity greater than or equal to 95%.
    \end{enumerate}
\end{itemize}

For example, the URL for a specific dataset could be: \url{https://qulacs-quantum-datasets.s3.us-west-1.amazonaws.com/test_mnist_784_f90.zip}.

\begin{figure}[H]
    \centering
    \includegraphics[width=0.8\linewidth]{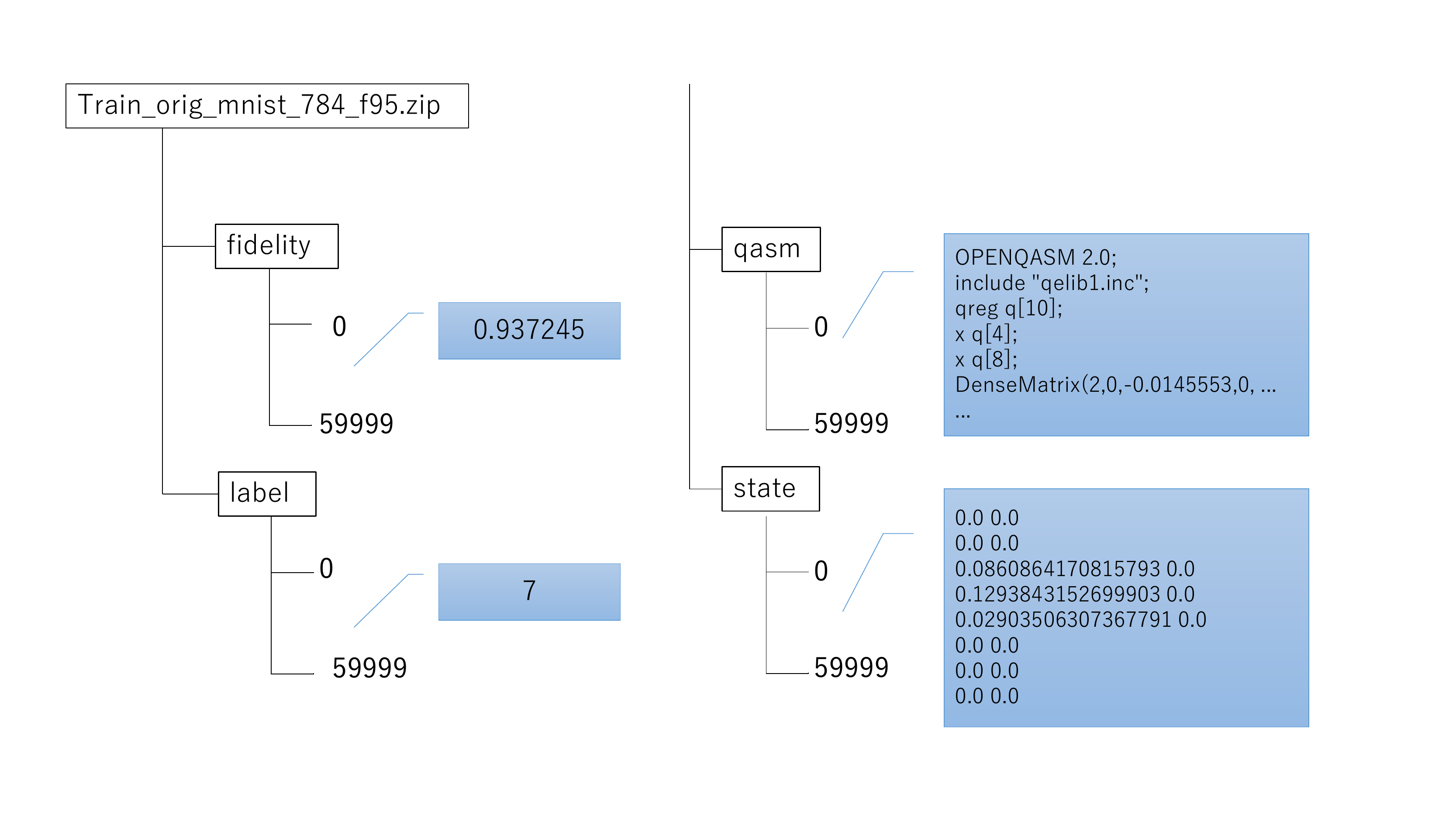}
    \caption{Example of a dataset directory structure that includes information about the fidelity, label, QASM files, and state vector for each image encoded in MNISQ.}
    \label{fig:DirectoryTree}
\end{figure}

As shown in Figure \ref{fig:DirectoryTree}, each dataset is represented as a dictionary with accessible key information. Although there are a total of nine datasets (three for each different fidelity level across the three datasets), options for augmentation or QASM formalism are available.
The dataset is composed of 9 subdatasets (60,000 training data - 10,000 test data) in two QASM formalism (but they are the same data) and 9 subdatasets from augmented data (480,000 training data each). The total is thus 4,950,000 data points.

\textbf{The official GitHub repository for the dataset is:} \url{https://github.com/FujiiLabCollaboration/MNISQ-quantum-circuit-dataset}.

%% file: arXiv_sections/QML.tex
\subsection{Quantum Machine Learning experiments}
\label{sec:qml}
In this section, we present the first quantum machine learning experiments on MNISQ. Our excellent results, when compared with the classical machine learning benchmark of the next section, show that our dataset may be quantum advantageous and represent the best performance in mnist dataset classification with any quantum machine learning method.

\begin{table}[htbp]
  \centering
  \caption{Experiments with the Quantum Kernel approach}
  \label{tbl:qml-benchmark}
  \begin{tabular}{cccc}
    \toprule
    \textbf{Dataset} & \textbf{Fidelity} & \textbf{1-versus-1 QSVM} & \textbf{1-versus-the-rest QSVM} \\
    \midrule
    {MNIST-784} & 80 & 0.9486 & 0.9494 \\
                               & 90 & 0.9733 & 0.9713 \\
                               & 95 & \textbf{0.9791} & 0.9777 \\
    \midrule
    {Fashion-MNIST} & 80 & 0.8204 & 0.8151 \\
                                   & 90 & 0.8541 & 0.8463 \\
                                   & 95 & \textbf{0.8678} & 0.8656 \\
    \midrule
    {Kuzushiji-MNIST} & 80 & 0.8417 & 0.8360 \\
                                     & 90 & 0.8909 & 0.8847 \\
                                     & 95 & \textbf{0.9066} & 0.9004 \\
    \bottomrule
  \end{tabular}
\end{table}

In table~\ref{tbl:qml-benchmark} are shown our results on the three subdatasets of MNISQ (mnist\_784, Fashion-MNIST, Kuzushiji-MNIST). The two methods employed are a variation of the quantum support vector machines (QSVMs) where, instead of performing a binary classification task, we perform a multiclass classification. We leave in the supplementary material an introduction to the classical SVMs, which differ from QSVM only in the computation of the kernel while leaving the same optimization procedure. The use of a quantum kernel is motivated by its ability to compute classically hard or intractable kernels~\cite{Mengoni2019KernelMI}. Recent studies have shown that quantum kernels can offer a quantum advantage over classical methods on carefully engineered datasets~\cite{liu2021rigorous}.

Looking at the table~\ref{tbl:qml-benchmark}, the \textit{one-versus-one} approach fits $n_{classes}(n_{classes}-1)$ QSVMs on all the possible couples of classes and finally classifies based on a voting scheme called \textit{one-versus-one}. On the other side, \textit{one-versus-the-rest} approach consists in training $n_{classes}$ QSVMs (10 for MNISQ subdatasets), each one able to classify one class versus any other. We performed our experiments using qulacs~\cite{qulacs} quantum computer simulator for the quantum kernels and sklearn implementations for the classifiers~\cite{scikit-learn}. The \textit{one-versus-the-rest} training is significantly faster because of its limited number of classifiers~\cite{bishop}, but since every classifier is trained on a small subset of the dataset, the \textit{one-versus-one} is likely more robust and, in fact, presents a slightly better performance. 
QSVMs differ from SVMs only because in the formulation of the decision function:
\begin{equation}
\label{dec-fun-kernel}
    f(\vec{x}) = {\rm sgn}\Bigl(\sum_{n=1}^N t_n \alpha_n K(\vec{x}^{(i)}, \vec{x})\Bigl), \vec{x}\in {\rm TestSet},
\end{equation}
The Kernel function $K(\vec{x}^{(i)}, \vec{x}^{(j)})$ is obtained by taking the inner product between quantum states. In our examples, the quantum states come from the encoding of our information (AQCE) available in the dataset, thus we can also write our quantum kernel as:
\begin{equation}
    \label{qk-def}
    K(\vec{x}^{(i)}, \vec{x}^{(j)}) = |\bra{00...0}AQCE(\vec{x}^{(i)})AQCE(\vec{x}^{(j)})^\dagger\ket{00...0}|^2
\end{equation}
\begin{figure}[H]
  \centering
  \includegraphics[width=0.8\linewidth]{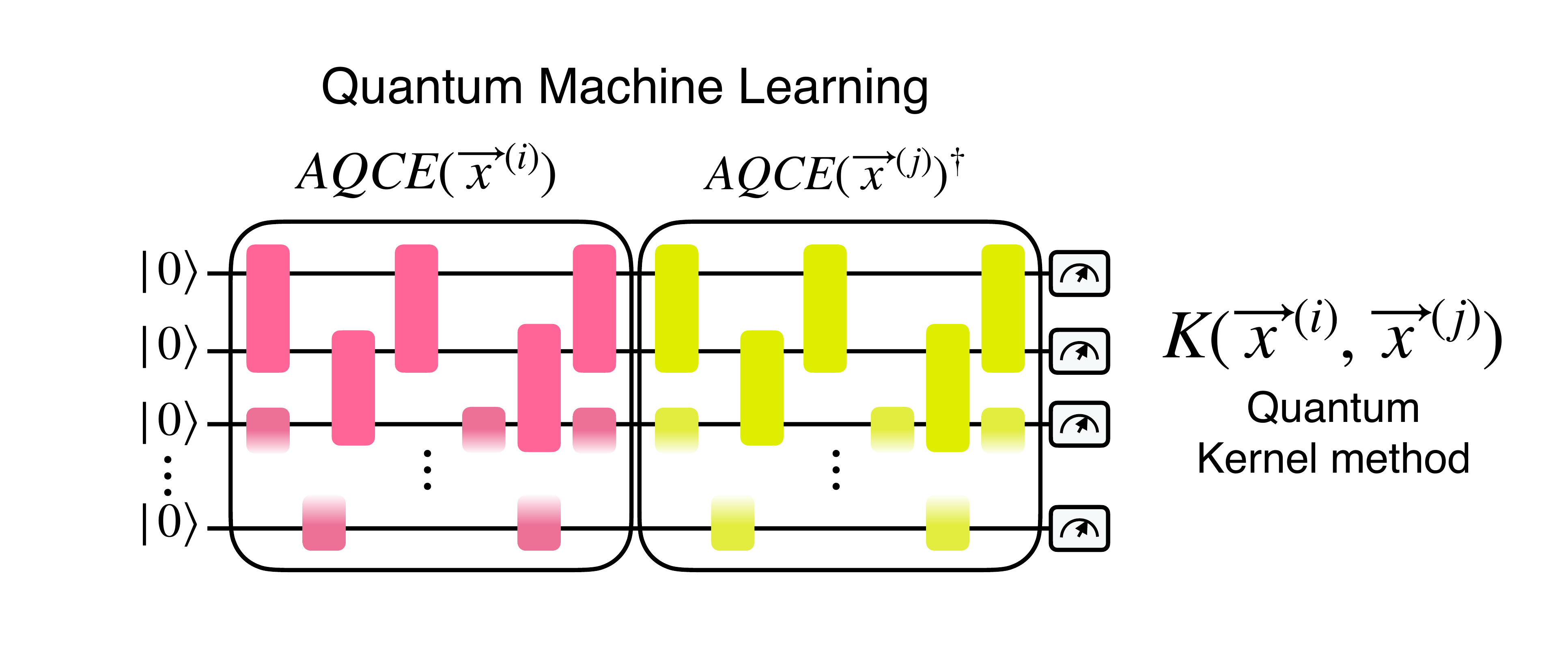}
  \caption{Example of the standard computation of the quantum kernel elements by generating the circuits $AQCE(\vec{x}^{(1)})AQCE(\vec{x}^{(2)})^\dagger$ for every pair of elements and computing the probability of the zero state after measurement in the computational basis.}
\end{figure}

We compare our prediction results based on the accuracy of the decision function, and we observe that our best prediction is reached when the fidelity is at least $95\%$ for the data. Thus, a higher fidelity circuit leads to a better approximation of the target state and a more effective feature map for the quantum kernel.

%% file: arXiv_sections/ML.tex
\subsection{Classical Machine Learning experiments}
\label{sec:ml}
\begin{figure}[htbp]
  \centering
  \includegraphics[width=0.8\linewidth]{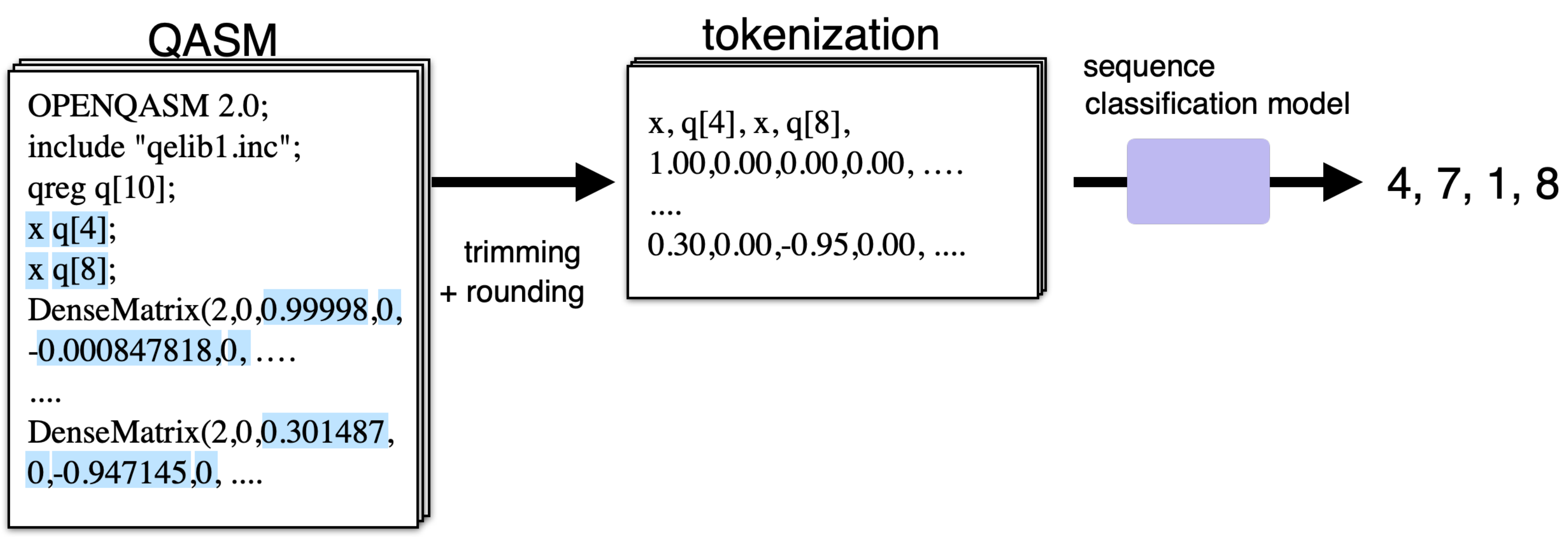}
  \caption{A schematic view of data processing in classical machine learning models.}
  \label{fig:schematic_classical_ml}
\end{figure}

We evaluate the performance of classical machine learning methods in recognizing quantum circuits by using classical deep neural networks for sequence classification. Specifically, we used Transformer~\cite{vaswani2017attention}, LSTM~\cite{hochreiter1997long}, and S4~\cite{gu2022efficiently} models as sequence classifiers, which are trained to classify QASMs (Quantum Assembly Language) in the MNISQ datasets.
Detailed experimental settings are described in the supplementary material.

The input QASMs were processed by tokenizing the text files, removing unnecessary information (such as headers), and rounding dense matrix elements (e.g., from $0.12345$ to $0.12$) as shown in Figure \ref{fig:schematic_classical_ml}. We specify that we performed the experiments using the version of the dataset with QASM files with the proprietary extension $\mathit{Dense()}$. The alternative QASM files, where $\mathit{Dense()}$ operators are substituted by gates descriptions, remain untested.

Table \ref{tab:classical_ml} presents test accuracy of the classical machine learning models on the MNISQ datasets. Among the models, S4 achieved the highest accuracy, with a test accuracy of $77.78\%$ on mnist\_784 with a fidelity of $95$. This performance surpassed that of the other models by a significant margin. Interestingly, as the fidelity decreased, both the accuracy of the quantum methods and the accuracy of the S4 and LSTM also dropped. On the other hand, the accuracy of the Transformer showed the opposite trend, where datasets with lower fidelity resulted in better performance.
Because higher-fidelity datasets have more information and consist of longer QASM sequences, these results suggest that S4 and LSTM can effectively capture long-range dependencies while Transformer cannot.
In particular, the S4 model was designed to capture long-range dependencies~\cite{gu2022efficiently}, and thus, achieved the highest performance.

\begin{table}[H]
  \centering
  \caption{Test accuracy of classical machine learning models.}
  \label{tab:classical_ml}
  \begin{tabular}{ccccc}
    \toprule
    \textbf{Dataset} & \textbf{Fidelity} & \textbf{S4} & \textbf{Transformer} & \textbf{LSTM} \\
    \midrule
    {MNIST-784} 
        & 80 & 66.61 & 43.42 & 59.06 \\
        & 90 & 72.41 & 41.85 & 60.88     \\
        & 95 & \textbf{77.78} & 40.40 & 66.93 \\
    \midrule
    {Fashion-MNIST} 
        & 80 & 64.93 & 46.22 & 59.30 \\
        & 90 & 67.91 & 43.71 & 63.19 \\
        & 95 & \textbf{71.42} & 43.60 & 64.64 \\
    \midrule
    {Kuzushiji-MNIST} 
        & 80 & 42.27 & 27.66 & 32.31 \\
        & 90 & 46.64 & 26.87 & 37.79 \\
        & 95 & \textbf{51.83} & 26.63 & 40.07 \\
    \bottomrule
  \end{tabular}
\end{table}

Table \ref{tab:classical_ml_augmented} shows test accuracy of the S4 model trained with and without augmented datasets.
We can observe clear performance improvement on each dataset, indicating the effectiveness of data augmentation in MNISQ as other machine learning problems.
We may also expect that dataset size matters to machine learning methods for quantum-circuit recognition tasks from these results.

\begin{table}[H]
  \centering
  \caption{Test accuracy of the S4 model trained with and without augmented data with a fidelity of 95.}
  \label{tab:classical_ml_augmented}
  \begin{tabular}{cccc}
    \toprule
    \textbf{Dataset} & \textbf{mnist\_784}   & \textbf{Fashion-MNIST}  &  \textbf{Kuzushiji-MNIST} \\
    \cmidrule{2-4}
    w/\quad  augmentation &  \textbf{81.36}  & \textbf{75.49}       & \textbf{57.67}    \\
    w/o augmentation & 77.78  & 71.42       & 51.83    \\
    \bottomrule
  \end{tabular}
\end{table}

It is important to note that these results were achieved without incorporating any prior knowledge of quantum computing into the classical machine learning methods. It is possible that incorporating such knowledge could further improve the performance of these classical models in classifying quantum circuits.

%% file: arXiv_sections/conclusion.tex
\section{Conclusion}
\label{sec:conclusion}
We have accomplished the generation of MNISQ, a comprehensive large-scale quantum circuit dataset comprising hundreds of thousands of samples. This task was achieved by embedding classical data into the complex amplitudes of the quantum state through AQCE. Given the critical role of datasets in advancing machine learning, MNISQ is poised to emerge as a standard quantum circuit dataset for quantum machine learning research.  Its importance lies in its ability to shape and advance the development of new techniques and approaches in this area of research. 
Our numerical experiments show that quantum machine learning show a high classification performance of over $97\%$. This result implicates that if we have a circuit that successfully embeds classical data into a quantum state, we can use quantum machine learning to achieve comparable high performance as classical machine learning. Hence, should the benefits of quantum machine learning be discovered, it may be prudent to consider storing the data in a condensed format, such as compressed quantum circuit data with preprocessing, offering the potential to enhance efficiency and optimize the utilization of quantum computing resources.

Furthermore, we have transcended the boundaries of quantum machine learning by applying classical machine learning models to our generated large-scale quantum circuit datasets. The remarkable outcome of this endeavor is that these classical models, devoid of any knowledge about the underlying mathematics of quantum mechanics and quantum computation within the quantum circuits, have achieved an astonishingly promising classification performance of up to $77\%$. This revelation underscores the incredible capability of classical machine learning in discerning the nature of computations performed by quantum circuits: an inherently complex and nontrivial task. Thus, our findings not only demonstrate the profound potential of quantum machine learning in realizing quantum advantages, but also unveil a groundbreaking capability of classical machine learning in comprehending the functionality of quantum circuits. The former represents a central challenge in the burgeoning field of quantum machine learning, while the latter has only become possible due to the advent of large-scale quantum circuit datasets. While constructing and optimizing quantum circuits remains a paramount task for harnessing the full potential of quantum computers, our study highlights the remarkable synergy between classical machine learning and this endeavor.

%% file: arXiv_supplementary/appendix.tex
\section{Supplements for the paper}
\label{sec:appendix}
\subsection{Related works}
Given that the field of defining datasets related to quantum information and quantum computing is still in its infancy, there is not an abundance of existing works in this direction. 
 Below we will introduce several existing studies concerning data associated with quantum phenomena, referred to as quantum datasets.

 {\bf NTangle~\cite{schatzki2021entangled}}:
 NTangle shares common ground with our research in the sense that it emphasizes the importance of focusing on quantum-related datasets to gain an advantage in QML. 
 However, it defines the quantum state itself as the dataset, which, unlike our quantum circuit dataset, cannot be directly applied to existing classical machine learning. Furthermore, it uses the property of the quantum state, whether it is entangled or not, as a label for the state. A distinctive feature of our approach is that labels are assigned based on the computational task that the quantum circuit is executing.

 {\bf QDataSet~\citesupp{perrier2022qdataset}}:
 QDataSet is a dataset defined from 52 types of data concerning quantum operations for single and two-qubit systems, encompassing both noise-inclusive and noise-free scenarios. 
 It is constructed to apply existing machine learning frameworks for characterizing quantum systems and improving experimental control setups including classical postprocesssing, and hence 
 QDataSet constitutes a dataset pertaining to the physical dynamics of one- or two-qubit systems.
 In our approach, we aim to explore machine learning using quantum computers and to apply existing machine learning frameworks to recognize the task done by quantum circuits,
 and hence we generate a large-scale quantum circuit dataset using quantum gates as fundamental units, avoiding elements considering the control of experimental systems, such as noise.

 {\bf Quantum datasets on Pennylane}:
 PennyLane is an open-source software library that provides tools for quantum machine learning and quantum computing. 
 Quantum Datasets in PennyLane, specifically for quantum chemistry and quantum spin systems provide detailed problem systems descriptions, parameterized models and their solution for these problems, aiding in the study of quantum algorithms related to these molecular and spin systems. 
 While these datasets serve as a valuable baseline for transparent research, how to apply quantum and classical machine learning for these data remain challenges.
 The size of the dataset is also too small to be applied for the state-of-the-art machine learning techniques.

 {\bf VQE-generated dataset~\cite{nakayama2023vqe}}:
The VQE-generated dataset is the most akin to our approach. It is a quantum circuit dataset labeled by the computational task it performs. 
However, a significant issue arises due to the dataset's size, as there are only 300 quantum circuits for each label, which is too small to apply state-of-the-art machine learning techniques. 
This problem is also the case for all above datasets,
which crucially causes lack of 
the state-of-the-art classical machine learning benchmarking.
To resolve this issue, we defined a large-scale quantum circuit dataset, MNISQ, applied conventional machine learning, and compared with QML.
\subsection{Basics of quantum computing}
In this section, we elucidate the fundamentals of quantum computation and quantum gates, 
introducing the notation employed in quantum information. 
Contrary to classical information, which is represented by binary bits $\{ 0,1\}$, 
quantum information defines the quantum bit, or qubit, as the smallest unit of information. 
This is represented as a linear combination of two orthogonal vectors, thus forming a superposition state:
\begin{align}
|\psi \rangle = \alpha |0\rangle + \beta |1\rangle.
\end{align}
We utilize Dirac's bra-ket notation to describe column vectors using $|\cdot \rangle$ symbol:
\begin{align}
|0\rangle =
\left( \begin{array}{c}
1
\\
0
\end{array}
\right),
\;\;\;
|1\rangle =
\left( \begin{array}{c}
0
\\
1
\end{array}
\right).
\end{align}
Furthermore, the coefficients of the linear combination, referred to as complex probability amplitudes, 
are normalized to satisfy $|\alpha|^2 + |\beta |^2 = 1$.
In addition to the column vector $|\psi \rangle$, we define the row vector obtained by taking the complex conjugate and transpose, or adjoint, of this:
\begin{align}
\langle \psi | = \alpha ^* \langle 0 | + \beta^* \langle 1 | = (|\psi \rangle )^{\dag}.
\end{align}
where
\begin{align}
\langle 0| = (|0\rangle )^{\dag}=(1,0), \;\;\;
\langle 1| = (|1\rangle )^{\dag}=(0,1).
\end{align}
Then, inner product of two vectors $|\psi \rangle , |\phi \rangle$ can be expressed as $\langle \phi | \psi \rangle$. Also, a linear operator can be represented by $|\phi \rangle \langle \psi |$.

For such a qubit, the operation executable on a quantum computer, the quantum gate, is specified by a $2\times2$ unitary matrix:
\begin{align}
\mathbf{U} =
\left( \begin{array}{cc}
u_{0|0} & u_{0|1}
\\
u_{1|0} & u_{1|1}
\end{array}
\right),
\end{align}
The reason for being limited to unitary matrices is to preserve the normalization condition. 
The symbol $|$ in-between the index is to explicitly show which left and right indices correspond to output and input, respectively.

When describing multiple qubits, we consider the multiple tensor product space of a qubit space as a unit. As a basis for the tensor product space by $n$ qubits, the direct product state of $\{|0\rangle, |1\rangle \}$ specified by the $n$-bit binary bit string $i_1 ... i_n$,
\begin{align}
|i_1 .... i_n \rangle = |i_1 \rangle \otimes \cdots \otimes | i_n\rangle
\end{align}
can be used. These $2^n$ orthonormal basis states are linearly combined to describe the quantum state of $n$ qubits using $2^n$ complex numbers:
\begin{align}
|\psi \rangle = \sum _{i_1 ... i_n} c_{i_1 ... i_n } |i_n\rangle 
\end{align}
Hence, the quantum state of $n$ qubits is represented as a $2^n$-dimensional complex vector. Similar to the case of a single qubit, the complex vector satisfies the normalization condition:
\begin{align}
\sum _{i_1 ... i_n} |c_{i_1 ... i_n } |^2 = 1.
\end{align}
When a quantum gate acts on the $k$-th qubit among $n$ qubits, its action is:
\begin{align}
U_{k} = I^{\otimes k-1} \otimes \mathbf{U} \otimes I^{\otimes (n-k)}
= \sum_{i_1 ... i_n } \sum_{j_k} u_{i_k| j_k} |i_1 ...i_k ... i_n \rangle \langle i_1 ...j_k i_{k+1} ... i_n|.
\end{align}
Universal quantum computation cannot be executed with operations acting only on one qubit. To construct a universal quantum computation, a two-qubit gate acting on two qubits is required. The two-qubit gate is specified by a $4\times 4$ matrix:
\begin{align}
\mathbf{W} =
\left( \begin{array}{cccc}
w_{00|00} & w_{00|01} & w_{00|10} & w_{00|11}
\\
w_{01|00} & w_{01|01} & w_{01|10} & w_{01|11}
\\
w_{10|00} & w_{10|01} & w_{10|10} & w_{10|11}
\\
w_{11|00} & w_{11|01} & w_{11|10} & w_{11|11}
\end{array}
\right).
\end{align}
When this two-qubit gate acts on the $k$-th and $l$-th ($l>k$) qubits, its action in the tensor product space of $n$ qubits is:
\begin{align}
W_{kl}
= \sum_{i_1 ... i_n } \sum_{j_k,j_l} u_{i_k i_l |j_k j_l}
|i_1 ...i_k ... i_l ... i_n \rangle \langle i_1 ...j_k i_{k+1} ... j_l i_{l+1} ... i_n|.
\end{align}
By appropriately applying the two-qubit gate, any unitary transformation can be constructed. It is also known that any two-qubit gate can be approximately constructed from the product of specific basic gates, $T$ gate, $H$ gate, CNOT gate, which are given by respectively:
\begin{align}
    H =
\frac{1}{\sqrt{2}} 
\left( \begin{array}{cc}
1 & 1 
\\
1 & -1
\end{array}
\right), \;\;\;
T=
\left( \begin{array}{cc}
e^{-i\pi/8} & 0
\\
0 & e^{i\pi/8}
\end{array}
\right), \;\;\;
{\rm CHOT} = 
\left( \begin{array}{cccc}
1 & 0& 0& 0
\\
0 & 1& 0& 0
\\
0 & 0& 0& 1
\\
0 & 0& 1& 0
\end{array}
\right).
\end{align}
However, in this study, we do not decompose the two-qubit gate into $\{ T, H , {\rm CNOT}\}$, and  we take the two-qubit gate as a basic unit.
A quantum algorithm starts from the quantum state of $n$ initialized qubits,
\begin{align}
|0^n \rangle =
\left( \begin{array}{c}
1
\\
\vdots
\\
0
\end{array}
\right)
\end{align}
and constructs a quantum circuit consisting of two-qubit gates according to the problem to be solved:
\begin{align}
\mathcal{C} = \prod _{m=1}^{M} W_{k_m l_m}^{(m)}
\end{align}
to obtain the output quantum state $|\psi \rangle = \mathcal{C} |0^n \rangle$. When a measurement is performed on this quantum state, a certain $n$-bit string $ b \in \{0,1\}^n$ is obtained with probability
\begin{align}
p_{b} = |\langle b | \psi \rangle| ^2
\end{align}
As will be explained later in the quantum kernel method, if there are two quantum circuits $\mathcal{C}$ and $\mathcal{C}'$, the square of the absolute value of the inner product between the quantum states $\mathcal{C}|0^n\rangle$ and $\mathcal{C}'|0^n \rangle$ generated by these can be written as
\begin{align}
p_{0...0} =| \langle 0 ^n | (\mathcal{C}')^{\dag} \mathcal{C} |0^n\rangle |^2 
\end{align}
From this, we can define a new quantum circuit $\mathcal{D} := (\mathcal{C}')^{\dag} \mathcal{C}$, and estimate the probability of obtaining all zeros to know the value of the inner product from the quantum computer.
%%%%%%%%%%%%%%%%%%%%%%%%%%%%%%%%%%%%
%%%%%%%%%%%%%%%%%%%%%%%%%%%%%%%%%%%%
\subsection{Automatic Quantum Circuit Encoding (AQCE) procedure}
We here give a detailed explanation of the AQCE algorithm.

Assuming that the quantum circuit $\mathcal{C}$ is composed of $M$ 2-qubit gates,
\begin{equation}
\mathcal{C} = \prod_{m=1}^M \mathcal{U}_m = \mathcal{U}_1 \mathcal{U}_2 \dots \mathcal{U}_M,
\end{equation}
With respect to the $m$th gate, $F$ can be rewritten as follows:
\begin{equation}
F_m = \bra{\Phi_{m-1}} \mathcal{U}_m^{\dagger} \ket{\Psi_{m+1}}.
\end{equation}
Here, the quantum states $\ket{\Phi_m}$ and $\ket{\Psi_m}$ are defined, respectively, by
\begin{align}
\ket{\Psi_m} &= \prod_{k=m}^M \mathcal{U}_k^{\dagger} \ket{\Psi} = \mathcal{U}_m^{\dagger} \mathcal{U}_{m+1}^{\dagger} \dots \mathcal{U}_M^{\dagger} \ket{\Psi} ,
\\
\bra{\Phi_m} &= \bra{0} \prod_{k=1}^m \mathcal{U}_k^{\dagger} = \bra{0} \mathcal{U}_1^{\dagger} \mathcal{U}_{2}^{\dagger} \dots \mathcal{U}_m^{\dagger}.
\end{align}
Then, in terms of trace formula, 
$F_m$ can be rewritten as
\begin{equation}
F_m = \text{Tr}[\ket{\Psi_{m+1}} \bra{\Phi_{m-1}} \mathcal{U}_m^{\dagger}] .
\end{equation}
If we denote $\mathbb{I}_m = {i,j}$ as the indices of the qubits on which $\mathcal{U}_m$ acts, 
we can introduce the fidelity tensor operator $\mathcal{F}_m$
\begin{equation}
\mathcal{F}_m = \text{Tr}_{{\mathbb{I}}_m}[\ket{\Psi_{m+1}} \bra{\Phi_{m-1}}],
\end{equation}
by which we can obtain $F_m$ as
\begin{equation}
F_m = \text{Tr}_{\mathbb{I}_m} [\mathcal{F}_m \mathcal{U}_m^{\dagger}],
\end{equation}
Furthermore, if we consider $\textbf{F}_m$ and $\textbf{U}_m$ as the matrix representations of $\mathcal{F}_m$ and $\mathcal{U}_m$, respectively, we obtain
\begin{equation}
F_m = \text{tr}[\textbf{F}_m \textbf{U}_m]
\end{equation}
The singular value decomposition of $\textbf{F}_m$ provides $\textbf{F}_m=\textbf{XDY}$,
where $\textbf{X}, \textbf{Y}$ are unitary matrices, and $\textbf{D}$ is a diagonal matrix with non-negative real diagonal elements $d_{n}$.
Then, we have
\begin{equation}
F_m = \text{tr}[\textbf{XDY} \textbf{U}_m^{\dagger}] = \text{tr}[\textbf{DZ}] = \sum_{n=0}^{2^2-1} d_n [\textbf{Z}]_{nn},
\end{equation}
where $\textbf{Z}=\textbf{Y}\textbf{U}_m^{\dagger} \textbf{X}$.
This leads 
\begin{equation}
|F_m| = \left| \sum_{k=0}^{2^2-1} d_n [\textbf{Z}]_{nn} \right| \leq \sum_{k=0}^{2^2-1} d_n \left| [\textbf{Z}]_{nn} \right|.
\end{equation}
Given that $\textbf{Z}$ is a unitary matrix, the absolute value of $F_m$ is maximized when $\textbf{Z}$ is an identity matrix. Therefore, the matrix $\textbf{U}_m$ that maximizes the absolute value of $F_m$ is given by
\begin{equation}
\textbf{U}_m = \textbf{XY}.
\end{equation}
Using this, the quantum circuit is optimized through the following steps:
\begin{enumerate}[(i)]
\item Set $m=1$.

\item Compute the representation matrix $\textbf{F}_m^{(k)}$ of the fidelity tensor $\mathcal{F}_m^{(k)} =\text{tr}_{{\mathbb{I}}_k}[\ket{\Psi_{m+1}} \bra{\Phi_{m-1}} \mathcal{U}_m]$ for all possible pairs $\mathbb{I}_k$ of indices.

\item Perform singular value decomposition $\textbf{F}_m^{(k)} = \textbf{X}^{(k)} \textbf{D}^{(k)} \textbf{Y}^{(k)}$ for all  $\textbf{F}_m^{(k)}$, and calculate $S^{(k)} = \sum_{n=0}^{3}[\textbf{D}^{(k)}]_{nn}$.

\item Find $k^{*}$ such that $S_k^{*}$ is maximized.

\item Set $\textbf{U}_m = \textbf{X}^{(k^{*})} \textbf{Y}^{(k^{*})}$, and apply the operator $\mathcal{U}_m$ specified by $\textbf{U}_m$ acting on $\mathbb{I}_{k^{*}}$.

\item If $m < M$, increment $m$ by 1 and return to step 2. If $m=M$, terminate the process.
\end{enumerate}
By following these steps, the structure of the quantum circuit, including which qubits the two-qubit gates should operate on, is optimized while autonomously constructing a quantum circuit that encodes the desired quantum state.
The fidelity of the embedded quantum state changes depending on the number of gates $M$ employed in the circuit $\mathcal{C}$.

The AQCE algorithm combines this circuit optimization step with a gate addition step to optimize the circuit while deepening it. In this work, we add new $\delta$ two-qubit gates until $M$ < $M_{\text{max}}$ or $F$ < $F_{\text{target}}$. Algorithm \ref{algorithm:AQCE} shows the AQCE algorithm procedure.

\begin{algorithm}
    \caption{AQCE}
    \label{algorithm:AQCE}
    \begin{algorithmic}[1]
        \renewcommand{\algorithmicrequire}{\textbf{Input:}}
        \renewcommand{\algorithmicensure}{\textbf{Output:}}
    \REQUIRE state $\ket{\Psi}$, integers $(M_0, \delta, M_{\text{max}}, N)$ and target fidelity $F_{\text{target}}$
    \ENSURE  circuit $\mathcal{C}$
    \\ \textit{Initialisation} : $\mathcal{C} \leftarrow I$ and $M = M_0$
    \WHILE{$M < M_{\text{max}} \ \land \ F < F_{\text{target}}$}
        \STATE add new $\delta$ two-qubit gates to $\mathcal{C}$
        \STATE $M \leftarrow M + \delta$
        \FOR{$counter = 1$ \algorithmicto \ $N$}
            \STATE optimize each of M gates (Above steps $(\mathrm{i}) \sim (\mathrm{vi})$)
        \ENDFOR
    \ENDWHILE
    \RETURN $\mathcal{C}$ 
    \end{algorithmic} 
\end{algorithm}
%%%%%%%%%%%%%%%%%%%%%%%%%%%%%%%%%%%%
\subsection{Support Vector Machines}
%%%%%%%%%%%%%%%%%%%%%%%%%%%%%%%%%%%%
Support Vector Machines (SVM) are supervised learning classifiers that learn a decision boundary between two classes of elements. We define the training data as $TR=\{\vec{x}^{(i)}, \vec{y}^{(i)}\}_{i=1:N} \in X\times Z_2$ and test data $TS=\{\vec{x}^{(i)}, \vec{y}^{(i)}\}_{i=1:M} \in X\times Z_2$, generally with $TS\cap TR = \emptyset$, where the two sets are assumed to have been generated by the same or highly similar distribution. SVMs solve a convex optimization problem on $TR$ to deliver a classifier for $TS$ with a trade-off between an accurate performance for the true labels and maximization of the orthogonal distance between the classes. The classification boundary is generally highly non-linear, therefore the data is first mapped with a feature function $\phi: X \rightarrow F$ to a higher dimensional space (feature space, in the quantum case the Hilbert Space), with a scalar product $\braket{\,,\,}$. In this space, the SVM is a linear classifier which aims at maximizing the margin between decision boundary and data points:
\begin{equation}
    \label{c-kernel:0}
    \arg\,\max_{w, b}\Biggl\{\frac{1}{||w||}\min_n[t_n(\braket{w,\phi(\vec{x}^{(n)})}+b)]\Biggl\}
\end{equation}
Here $t_n$ are target class values for the elements $\vec{x}^{(i)} \in TR$, the $\frac{1}{||w||}$ is just a scaling factor to avoid dependence on $n$. As an intuition, $\vec{y}(\vec{x}^{(n)}) = \braket{w,\phi(\vec{x}^{(n)})}+b$ defines the classification in one of the two classes based on its positive or negative sign. Instead, $t_n(\braket{w,\phi(\vec{x}^{(n)})}+b)$ is the distance from the decision surface of the problem (which then we want to maximize). In our case, we use a soft margin SVM, which means that $t_ny(\vec{x}^{(n)}) \geq 1-\xi_n$, where $\xi_n$ are slack variables that help to deal with overlapping class distributions giving a small penalty for misclassification. The previous formulation of \ref{c-kernel:0} is too complex to solve, therefore it is generally converted to the equivalent \textit{primal problem}:
\begin{mini}
    %\label{c-kernel:1}
    {w, b, \xi}{\frac{1}{2}||w||^2 + C\sum_{n=1}^N\xi_n}{}{}
    \addConstraint{t_n(\braket{w,\phi(\vec{x}^{(n)})}+b)}{\geq}{1-\xi_n}
    \addConstraint{\xi_n}{\geq}{0}
\end{mini}\\
The “Kernel trick” consists in shifting this formulation to one where we don't have to compute the computationally demanding mapping $\Phi$ for every point, but instead just compute a Kernel function defined as $K(\vec{x}^{(i)}, \vec{x}^{(j)}) = \braket{\phi(\vec{x}^{(i)}),\phi(\vec{x}^{(j)})}$.  Follows the dual formulation:
\begin{maxi}
    %\label{c-kernel:1}
    {\alpha_i \in R}{\sum_{n=1}^N\alpha_n-\frac{1}{2}\sum_{n=1}^N\sum_{m=1}^N\alpha_n \alpha_m t_nt_mK(\vec{x}^{(n)}, \vec{x}^{(m)})}{}{}
    \addConstraint{0 \leq \alpha_i}{\leq}{ C}
    \addConstraint{\sum_{m=1}^N\alpha_n t_n}{=}{0}
\end{maxi}\\
Solving the dual problem allows then to solve the primal and the relationships for the equivalence allows expressing the classification function as:
\begin{equation}
    f(\vec{x}) = sgn\Bigl(\sum_{n=1}^N t_n \alpha_n K(\vec{x}^{(i)}, \vec{x})\Bigl), \vec{x}\in TS 
\end{equation}
%%%%%%%%%%%%%%%%%%%%%%%%%%%%%%%%%%%%

%%%%%%%%%%%%%%%%%%%%%%%%%%%%%%%%%%%%

%% file: arXiv_supplementary/data_gen.tex
%%%%%%%%%%%%%%%%%%%%%%%%%%%%%%%%%%%%
\section{Details on Dataset Generation}

This section provides an in-depth look at the data and hardware used for dataset generation.

\begin{table}[h!]
  \caption{Parameters for Achieving Specified Fidelity}
  \centering
  \label{table:fidelity-parameters}
  \begin{tabular}{ccc}
    \toprule
    Fidelity (\%) & M (Max Number of 2-Qubit Gates) & $\delta$ (Incremental Gate Factor) \\
    \midrule
    80 &  25 &  3  \\
    90 &  50 &  6  \\
    95 &  100 &  12  \\
    \bottomrule
  \end{tabular}
\end{table}

Table \ref{table:fidelity-parameters} outlines the parameters used to ensure the attainment of the specified fidelity. Here, 'M' represents the maximum number of 2-qubit gates, while '$\delta$' is the factor determining the number of gates to be added in each iteration.

The dataset was generated using a cluster machine with the following specifications:

\begin{itemize}
    \item CPU: Dual Intel Xeon Platinum 9242 (2.3GHz, 48-Core)
    \item Memory: 384 GB (24 x 16 GB DDR4 -2933 Reg.ECC)
\end{itemize}

The generation of one MNIST image with $90\%$ fidelity took approximately 8 seconds per process. Although the AQCE algorithm can be run in multiple threads, we opted for single-threaded execution due to its efficiency when generating a large number of images in parallel.

We divided the generation of 480,000 data sheets into 1000 processes, with each process generating 480 sheets. Each process was registered as a single job on the cluster machine. Considering the three types of fidelity and the original 60,000 sheets of learning data, the number of jobs increased proportionally. To avoid disk IO contention, we stored 480 copies in memory and wrote them to the disk all at once after all 480 circuits were generated.

Our implementation uses Qulacs~\cite{qulacs} internally and runs within a container using Singularity to separate the cluster environment from the execution environment. Using a 40-node cluster (3840 cores), we were able to generate 4.95 million pieces of data in approximately 8.5 hours.

\begin{table}[h!]
  \caption{Data Breakdown for 4.95 Million Records}
  \label{table:data-breakdown}
  \centering
  \begin{tabular}{llll}
    \toprule
    \textbf{Dataset} & \textbf{Description} & \textbf{Metric} & \textbf{Augmented Training Data?} \\
    \midrule
    MNIST &  60,000 Training Data, 10,000 Test Data &  Fidelity  & Yes (480,000)  \\
    Fashion MNIST &  60,000 Training Data, 10,000 Test Data &  Fidelity & Yes (480,000)   \\
    Kuzushiji MNIST &  60,000 Training Data, 10,000 Test Data &  Fidelity  & Yes (480,000)  \\
    \bottomrule
  \end{tabular}
\end{table}

Table \ref{table:data-breakdown} provides a breakdown of the data for the 4.95 million records generated. Each dataset includes original training and test data, and is augmented with additional training data to enhance the learning process.

\subsection{Dual-Form Representation: Quantum Circuits and QASM Descriptions}

Quantum circuits for Qulacs can be generated with a specific dataset using the following methods:

\begin{verbatim}
qulacs_dataset.[type].loader.load_[type]_[data_type]_[fidelity]
\end{verbatim}

In the above method:

\begin{itemize}
    \item \textbf{type} refers to the type of dataset, which can be mnist, fashion\_mnist, or kuzushiji\_mnist.
    \item \textbf{data\_type} indicates whether the data is original, train, or test.
    \item \textbf{fidelity} represents the fidelity level, which can be f80, f90, or f95.
\end{itemize}

For instance, to load the MNIST training dataset with a fidelity of 95\%, you would use:

\begin{verbatim}
qulacs_dataset.mnist.loader.load_mnist_train_f95()
\end{verbatim}

Running a circuit embeds a dataset into a quantum state. This embedded quantum state can then be used to add a quantum machine learning circuit for learning purposes.

The quantum circuit fetches QASM data from the cloud and transforms it into a quantum circuit. The DenseMatrix feature is used for embedding the features of your dataset. 

Please note that \textit{DenseMatrix} is not a part of OpenQASM and is a proprietary extension with the following format:

\begin{itemize}
    \item \textbf{Format:}
\begin{verbatim}
DenseMatrix (number of target qubits, number of control qubits,
    matrix elements a + bi ordered by "a, b"),
    target qubit column, control qubit column;
\end{verbatim}

    \item \textbf{Example:}
\begin{verbatim}
DenseMatrix(2,0,0.900726,0,-0.267421,0, ... ,0.752815,0) q[2],q[6];
\end{verbatim}
\end{itemize}

The dataset directory structure is organized as follows:

\begin{itemize}
      \item \textbf{fidelity:} This represents the fidelity between the original true quantum state and the quantum state embedded by executing the quantum circuit.
      \item \textbf{label:} These are the labels for the dataset.
      \item \textbf{qasm:} These are the QASM files that are converted into quantum circuits.
      \item \textbf{state:} This is the value of the state vector that is embedded as a quantum state.
\end{itemize}
Note that the repository may be updated over time, so we will specify on the GitHub page any change in format or syntax.
%%%%%%%%%%%%%%%%%%%%%%%%%%%%%%%%%%%%

%% file: arXiv_supplementary/S4.tex
\subsection{Structured State Space Sequence model (S4)}

The Structured State Space sequence model (S4) is a neural network based on state space models designed to handle long sequences efficiently~\cite{gu2022efficiently}.
It  is the first model to solve the Path-X task in the Long Range Arena benchmarks~\citesupp{tay2021long} that requires the ability to handle complex long-range dependencies, which other previous methods including Transformer failed to learn.

%% file: arXiv_supplementary/TF.tex
\subsection{Transformer}

Transformer is a neural network model consisting of self-attention and fully-connected feed-forward network~\cite{vaswani2017attention}.
Although it was originally proposed for machine translation, Transformer and its variants are now popularly used in various tasks, including computer vision~\citesupp{dosovitskiy2021an}, and achieve state-of-the-art performance.
Compared with recurrent neural networks (RNNs) or convolutional neural networks (CNNs), Transformer depends on less inductive biases and thus expected to learn them from data.

%% file: arXiv_supplementary/LSTM.tex
\subsection{Long Short-Term Memory (LSTM)}

Long Short-Term Memory (LSTM) is a type of recurrent neural network (RNN) and has been the standard neural network for sequence modeling for a long time~\cite{hochreiter1997long}.
Compared with the vanilla RNN, LSTM has additional gates to control information so that it can handle longer sequences.
These gates avoid gradient vanishing or gradient explosion, which the vanilla model suffers from, enabling it to apply to complex sequence tasks, such as machine translation~\citesupp{sutskever2014sequence}.

%% file: arXiv_supplementary/models.tex
\subsection{Experimental settings and models}

In this section, we provide a comprehensive overview of the configuration details for the S4, Transformer, and LSTM models used in our experiments.
We trained these models on the proposed datasets in the QASM format for 200 epochs.

The input QASMs were preprocessed by removing superfluous information such as headers and rounding elements of dense matrices to \emph{one} decimal place, as illustrated in Figure 5 of the paper.
Interestingly, rounding to two decimal places decreased the performance.
Test accuracy of S4 dropped from 77.78\% to 72.10\% for mnist\_784 classification at a fidelity of 95\%.

\begin{verbatim}
    # The original format:
    DenseMatrix(2,0,0.541645,0,-0.038637,0, ... ,0.540171,0) q[0],q[1];

     # The preprocessed format:
     0.5, 0, 0, 0, ..., 0.5, q[0], q[1]
 \end{verbatim}
 
\subsubsection{S4 Model}

The S4 model was implemented using a JAX-based version available at \url{https://srush.github.io/annotated-s4/}. We adopted its “CIFAR-10 classification” configuration, modifying the number of epochs.

The hyperparameters used for training the S4 model are as follows:

\begin{itemize}
    \item \textbf{Hidden size}: 512
    \item \textbf{Number of layers}: 6
    \item \textbf{Dropout rate}: 0.25
    \item \textbf{Optimizer}: AdamW (initial learning rate of $1.0\times10^{-3}$ with cosine annealing with warmup and weight decay rate of $0.01$)
    \item \textbf{Batch size}: 50
\end{itemize}

\subsubsection{Transformer Model}

We adopted a BERT-like Transformer architecture as a sequence classifier~\citesupp{devlin2018bert} using the Transformer Encoder module in PyTorch. Layer normalization was applied prior to self-attention and feedforward layers (pre-norm), and the GELU activation function was used as \cite{vaswani2017attention}.

The following hyperparameters are used for training the Transformer models.

\begin{itemize}
    \item \textbf{Hidden size}: 512
    \item \textbf{Number of heads}: 8
    \item \textbf{Number of layers}: 6
    \item \textbf{Dropout rate}: 0.10
    \item \textbf{Optimizer}: AdamW (initial learning rate of $1.0\times10^{-5}$ with cosine annealing with warmup and weight decay rate of $0.01$)
    \item \textbf{Batch size}: 128
\end{itemize}

We selected the initial learning rate from $\{1.0\times10^{-5}, 5.0\times10^{-5}, 1.0\times10^{-4}\}$ and weight decay from $\{0.01, 0.05, 0.1\}$ on mnist\_786 with a fidelity of 80.

\subsubsection{LSTM Model}

We implemented the LSTM model using the LSTM module in PyTorch followed by a linear layer with the following hyperparameters.

\begin{itemize}
    \item \textbf{Hidden size}: 512
    \item \textbf{Number of layers}: 2
    \item \textbf{Dropout rate}: 0.25
    \item \textbf{Optimizer}: AdamW (initial learning rate of $1.0\times10^{-3}$ with cosine annealing with warmup and weight decay rate of $0.01$)
    \item \textbf{Batch size}: 128
\end{itemize}

We selected the initial learning rate from $\{1.0\times10^{-4}, 1.0\times10^{-3}\}$ on mnist\_786 with a fidelity of 80.

\subsection{Experiments with different data preprocessing}
We show here alternative experiments using a different data processing strategy from the QASM files testing on Transformer and LSTM.
The real part of the DenseMatrix in QASM is truncated to three decimal places and arranged in a specific format.
\begin{verbatim}
 # The original format:
 DenseMatrix(2,0,0.541645,0,-0.038637,0, ... ,0.540171,0) q[0],q[1];

 # is converted to:
 0.541 -0.038 ... 0.54 q[0] q[1] _
 \end{verbatim}

 The classification process involves the following steps:

 \begin{itemize}
     \item \textbf{Data Conversion:} If there are 25 DenseMatrix instances, they are juxtaposed to form a single sequence.
    
     \item \textbf{Vocabulary Building:} The vocabulary is built from the real part (truncated to three decimal places), $q[0]$, and the data delimiter, $\_$. The following command can be used to build the vocabulary:
     \begin{verbatim}
     onmt_build_vocab -config config.yaml -n_sample 1000
     \end{verbatim}
    
     \item \textbf{Training:} The model can be trained using the following command:
     \begin{verbatim}
     onmt_train -config config.yaml
     \end{verbatim}
 \end{itemize}

\begin{table}[H]
  \centering
  \caption{Test accuracy of Transformer and LSTM with different preprocessing from before.}
  \label{tab:classical_ml2}
  \begin{tabular}{cccc}
    \toprule
    \textbf{Dataset} & \textbf{Fidelity} & \textbf{Transformer} & \textbf{LSTM} \\
    \midrule
    MNIST-784 
        & 80  & \textbf{58.84} & 55.68 \\
        & 90  & \textbf{57.74} & 55.14 \\
        & 95  & 54.90 & \textbf{55.68} \\
    \midrule
    Fashion-MNIST 
        & 80  & \textbf{62.70} & 55.00 \\
        & 90  & \textbf{57.96} & 55.54 \\
        & 95  & 55.00 & 55.00 \\
    \midrule
    Kuzushiji-MNIST 
        & 80  & \textbf{56.39} & 55.00 \\
        & 90  & \textbf{55.12} & 55.11 \\
        & 95  & \textbf{55.12} & 55.00 \\
    \bottomrule
  \end{tabular}
\end{table}

\subsubsection{relative Transformer Model}

For the Transformer and LSTM models, we employed OpenNMT-py \citesupp{opennmt}. The specific configuration for the Transformer model is detailed below:

\begin{itemize}
    \item \textbf{Word vector size}: 512
    \item \textbf{Hidden size}: 512
    \item \textbf{Number of layers}: 6
    \item \textbf{Feed-forward size (transformer\_ff)}: 2048
    \item \textbf{Number of heads}: 8
    \item \textbf{Accumulation count (accum\_count)}: 8
    \item \textbf{Optimizer}: Adam (Beta1: 0.9, Beta2: 0.998)
    \item \textbf{Decay method}: Noam
    \item \textbf{Learning rate}: 2.0
    \item \textbf{Maximum gradient norm (max\_grad\_norm)}: 0.0
    \item \textbf{Dropout rate}: 0.1
    \item \textbf{Label smoothing}: 0.1
\end{itemize}

\subsubsection{relative LSTM Model}

The LSTM model was constructed using the following components, each with their respective hyperparameters:

\begin{itemize}
    \item \textbf{RNNEncoder}: LSTM encoder with 500 hidden units
    \item \textbf{RNNDecoder}: Stacked LSTM decoder with two LSTM cells (First cell: 1000 hidden units, Second cell: 500 hidden units)
    \item \textbf{GlobalAttention}: Implemented in the model
    \item \textbf{Generator}: Linear layer with 500 units for output generation
    \item \textbf{Word vector size}: 500
    \item \textbf{Hidden size}: 500
    \item \textbf{Number of layers}: 2
    \item \textbf{Optimizer}: SGD
    \item \textbf{Accumulation count}: 1
    \item \textbf{Learning rate}: 1.0
    \item \textbf{Maximum gradient norm}: 5
    \item \textbf{Dropout rate}: 0.3
\end{itemize}

%% file: arXiv_supplementary/license.tex
\section{Dataset license}
\label{sec:license}
We release MNISQ dataset with the following license:

\textbf{CC BY-SA 4.0}

\url{https://creativecommons.org/licenses/by-sa/4.0/}

%% file: arXiv_supplementary/metadata.tex
\section{Metadata}
We added structured metadata to the dataset's metadata page. We are also including the metadata on the GitHub repository of the dataset \url{https://github.com/FujiiLabCollaboration/MNISQ-quantum-circuit-dataset/blob/main/metadata}.

%% file: arXiv_supplementary/statement.tex
\section{Authors statement}
\label{sec:statement}
As authors of the MNISQ dataset, we affirm that we bear all responsibility in case of any violation of rights or any legal issues that may arise from the use of the dataset. We also confirm that the MNISQ dataset is licensed under the Creative Commons Attribution-ShareAlike 4.0 International (CC BY-SA 4.0) license, allowing users to share and adapt the dataset as long as proper attribution is provided, and any derivative works are shared under the same license.

%% file: arXiv_supplementary/url.tex
\section{GitHub url}
The dataset and the experiments are accessible on the GitHub URL: \url{https://github.com/FujiiLabCollaboration/MNISQ-quantum-circuit-dataset}. We will keep working on improving the directory, but the dataset is already accessible as specified in the paper and there are a few tutorials. We will maintain this repository and provide updates on the dataset.